\documentclass[twocolumn,showpacs,preprintnumbers,amsmath,amssymb]{revtex4}

\usepackage{graphicx}
\usepackage{dcolumn}
\usepackage{bm}

\begin{document}

\preprint{APS/123-QED}

\title{Hyperfine anomaly in Be isotopes and the neutron
spatial distribution; a three-cluster model for $^9$Be.}

\author{Y. L. Parfenova}
\email{Yulia.Parfenova@ulb.ac.be}
\altaffiliation[Also at ]{ Skobeltsyn Institute of Nuclear Physics, Moscow State
University, 119992 Moscow, Russia.}
\affiliation{%
Physique Nucl\'eaire Th\'eorique et 
Physique Math\'ematique, CP229, Universit\'e Libre de Bruxelles
B 1050 Brussels, Belgium.}
\author{Ch. Leclercq-Willain}
\email{cwillain@ulb.ac.be}
\affiliation{%
Physique Nucl\'eaire Th\'eorique et 
Physique Math\'ematique, CP229, Universit\'e Libre de Bruxelles
B 1050 Brussels, Belgium.}


\date{\today}

\begin{abstract}
The study of the hyperfine (hfs) anomaly in neutron rich nuclei 
can give a very specific and unique way to study the neutron
distribution and the clustering structure. We study the 
sensitivity of the hfs anomaly to the clustering effects 
in the $^9$Be isotope using two
different nuclear wave functions obtained in the three-cluster 
($\alpha+\alpha+n$) model. 
The results are compared to those obtained for $^{9,11}$Be in 
a two-body 
core + neutron model to examine whether the hfs anomaly 
is sensitive to a halo structure in $^{11}$Be.
\end{abstract}

\maketitle

\section{Introduction \label{sec:s1}}

The study of the hyperfine (hfs) anomaly in neutron rich nuclei,
in particular, those with loosely bound neutrons
can give a very specific 
and unique way to measure the neutron distribution. In a previous paper \cite{Par05} we have obtained the values of
the hfs anomaly calculated in a two-body core + neutron model 
for the $^{9,11}$Be isotopes.
The hfs anomaly $\epsilon$ is defined as the sum of the hfs anomaly 
related to the Bohr-Weisskopf effect ($\epsilon_{BW}$)\cite{Boh50}
and of the Breit-Rosenthal-Crawford-Shawlow (BRCS) correction
$\delta$ \cite{Ros72}, $\epsilon=\epsilon_{BW}+\delta$. 

It was found in Ref.\cite{Par05} that in  
Be isotopes, the 
value $\epsilon_{BW}$ is comparable to the BRCS correction 
$\delta$. The value $\epsilon_{BW}$ is very sensitive to
the weights of the partial states in the ground 
state wave function 
and might vary within 50\% depending on those weights. 
In \cite{Par05}, we found a difference of about 25\% for the total hfs 
anomaly value in the $^{9,11}$Be isotopes.

In the present paper, we calculate the hfs anomaly in  
a three-cluster model of $^9$Be and study the sensitivity of 
the hfs anomaly to the clusterization effects. The calculations are performed with two different three-body wave 
functions \cite{Priv05,Gri02}. These wave functions differ from each other by the choice
of the cluster-cluster interaction potentials used in their calculations.
They are characterized by the same partial states but contributing with different weights.
They both reproduce the $^{9}$Be 
rms radius and the three-cluster ($\alpha$+$\alpha$+$n$) dissociation energy
(1.573 MeV).
We compare the results obtained with these two wave functions, and 
also compare the present results with the core + neutron model
calculations \cite{Par05,Fuj99}.

The accuracy of the description of the $^{9}$Be ground state wave function can be explored
by calculating the magnetic dipole and electric quadrupole moments which are
determined by the weights and quantum numbers of the states. 
The $^{9}$Be magnetic dipole moment is independent 
of the radial behavior of the wave function but this is not 
the case for the electric quadrupole  moment.
Calculation of both the magnetic dipole and electric quadrupole moments thus provides a 
precise test of the accuracy of all aspects of the ground state wave function (for the 
experimental data on the magnetic dipole moment in the Be isotopes, we
refer to Refs. \cite{Web76}, \cite{Dic49}; no experimental data exists on the hyperfine anomaly 
for these nuclei). We also compare the results for $^{9}$Be$^+$ to that for
$^{11}$Be$^+$, to investigate the sensitivity of the hfs anomaly to diffuse (halo)neutron structures.



\section{Three-body wave function \label{sec:s3}}

The wave 
function $\Psi _{JM}^{3b}$ of the fragment relative motion  
in the three-cluster model ($\alpha +\alpha +n$) of $^{9}$Be 
is described with the Jacobi coordinates in the 
method of hyperspherical harmonics \cite{Zhu93,Efr72}. 
It is written as
\[
\Psi _{JM}^{3b}({\bf \xi_1},{\bf \xi_2})=
\sum \limits_{KLSS_x l_{1}l_{2}M_{L}} 
[\digamma _{LM_{L}}^{KSS_{x}l_{1}l_{2}} \otimes 
\chi_{SM-M_{L}}^{s_1(s_2s_3)S_x}]^{JM},
\]
where 
$\digamma _{LM_{L}}^{KSS_{x}l_{1}l_{2}}$ is the "active part" 
of the three-body wave function carrying the total orbital angular 
momentum $L$ with the projection $M_L$. $\chi _{SM_{S}}$ is 
the total spin function of the whole system with the total spin $S$ and 
projection $M_{S}$ (here restricted to the neutron spin function).
 The wave function 
$\digamma _{LM_{L}}^{KSS_{x}l_{1}l_{2}}(\xi_1 ,\xi_2 )$ depends 
on the relative coordinates (hyperradius and relative angles) 
and the neutron spin 
\[
\digamma _{LM_{L}}^{KSS_{x}l_{1}l_{2}}(\xi_1 ,\xi_2 )=\rho ^{-5/2} \Re
_{Kl_{1}l_{2}}^{LSS_x}(\rho )\Im _{KLM_{L}}^{l_{1}l_{2}}(\Omega _{5})  , 
\]
where $l_{1}$ and $l_{2}$ are the angular momenta of the relative 
motion corresponding to the $\xi_1$ and $\xi_2$ coordinates. 
The sum $\vec{L}=\vec{l}_{1}+\vec{l}_{2}$ gives
the total angular momentum; $K$ is the hypermomentum. 

The hyperradius $\rho ^{2}=\xi_1^{2}+\xi_2^{2}$ is a collective 
rotationally and permutationally invariant variable, $\xi_1$ and 
$\xi_2$ are the translationally invariant normalized sets of 
Jacobi coordinates.
We choose the Jacobi coordinates as follows:
\begin{eqnarray}
{\bf x} &=&(A_{23})^{1/2}{\bf{r_{23}}}, \label{jacxy} \\
{\bf y} &=&(A_{1(23)})^{1/2}{\bf{r_{1(23)}}}, \nonumber
\end{eqnarray}
where
${\bf{r_{23}}}$ is the relative coordinate of fragments 2 and 3, 
and ${\bf{r_{1(23)}}}$ is the coordinate of fragment 1 relative 
to the center of mass of the fragments 2 and 3. $A_{23}$ is the 
reduced mass number for the pair (2,3),
similarly, $A_{1(23)}$ is the reduced mass of the fragment 1 with respect to the mass of the 
subsystem $(2,3)$.
 

The five hyperspherical polar angles are 
$\Omega _{5}=\{\theta ,\hat{x},\hat{y}\}$ where 
$\theta$ is defined by the relations
\begin{eqnarray}
\xi_1 \equiv{x} &=&\rho \sin \theta, \label{coorch} \\
\xi_2 \equiv{y} &=&\rho \cos \theta. \nonumber
\end{eqnarray}
The choice of three different systems of Jacobi coordinates 
leads to three-body wave functions with different phase factor, 
the three different Jacobi coordinates systems being connected 
together by defined rotations. 
For $^{9}$Be, the T-basis correspond to choose the neutron as
fragment 1, the two $\alpha$-particles as fragments 2 and 3. 
The Y-basis associates the fragment 1 with one of the 
$\alpha$-particles.

The values of the hypermomentum are $K=l_{x}+l_{y}+2n $ ($n=1,2,...$).
The hyperspherical harmonics have the form
\[
\Im _{KLM_{L}}^{l_{x}l_{y}}(\Omega _{5})=\psi _{K}^{l_{x}l_{y}}(\theta
)[Y_{l_{x}}(\hat{x}) \otimes Y_{l_{y}}(\hat{y})]_{LM_{L}}, 
\]
where 
\[
\psi _{K}^{l_{x}l_{y}}(\theta )=N_{K}^{l_{x}l_{y}}(\sin \theta
)^{l_{x}}(\cos \theta )^{l_{y}}P_{n}^{l_{x}+1/2,l_{y}+1/2}(\cos 2\theta ) 
\]
and $P_{n}^{\alpha ,\beta }$ is a Jacobi polynomial. 
$N_{K}^{l_{x}l_{y}}$ is the coefficient of normalization
\[
N_{K}^{l_{x}l_{y}}=\sqrt{\frac{2(n!)(K+2)(n+l_{x}+l_{y}+1)!}{\Gamma
(n+l_{x}+3/2)\Gamma (n+l_{y}+3/2)}} 
\]
and the normalization condition for the function 
$\psi _{K}^{l_{x}l_{y}}(\theta) $ is 
\[
\int\limits_{0}^{\pi /2}\psi _{K}^{l_{x}l_{y}}(\theta )\psi _{K^{\prime
}}^{l_{x}^{\prime }l_{y}^{\prime }}(\theta )\sin ^{2}\theta \cos ^{2}\theta
d\theta =\delta _{KK^{\prime }}\delta _{l_{x}l_{x}^{\prime }}\delta
_{l_{y}l_{y}^{\prime }} . 
\]

The charge density distribution of the three-body nucleus 
entering the calculation of the electronic wave functions 
can be obtained as 
\begin{equation}
\rho(\bf r)=\rho _{0}<\Psi _{JM}^{3b}\left| 
\sum \limits_{i} Z_{i}\rho _{i}({\bf r},{\bf x},{\bf y})\right| 
\Psi _{JM}^{3b}> , \label{rho}
\end{equation}
where $\rho_{0}$ is a normalization factor. In the case of
$^{9}$Be, the $\alpha $-particle density distribution
is approximated by a sum of 
Gaussians, with the parameters taken from \cite{deV87} and giving a charge radius
equal to 1.676 fm.

In the proceeding calculations two differing evaluations of $\Psi _{JM}^{3b}$ \cite{Priv05,Gri02} as noted in Section \ref{sec:s1} are used. 

\section{Magnetic hyperfine structure \label{HFSsec}}

Here, we briefly mention the main points of the formalism
for the hfs anomaly calculations. 
For more details we refer to \cite{Par05} and references therein.

The magnetic hyperfine interaction Hamiltonian is defined by 
\begin{equation}
\mathcal{H}=-\int \bf{\mathcal{J}}(\mathbf{r})\cdot \mathbf{A}(\mathbf{r})\,d^{3}r,
\end{equation}
where $\bf{\mathcal{J}}$ is the nuclear current density and 
$\mathbf{A}$ is the
vector potential created by the atomic electrons.

The hyperfine interaction couples the electronic angular momentum 
$\mathbf{J}$ and the nuclear one $\mathbf{I}$ to a hyperfine 
momentum $\mathbf{F} = \mathbf{J} + \mathbf{I}$. The magnetic 
hyperfine splitting energy $W$ for a state $\mid I J F M_F=F >$ 
is defined as the matrix element of the Hamiltonian $\mathcal{H}$.

The functions $F^{\kappa J}$, $G^{\kappa J}$ entering the matrix
element are the radial parts
of the large and small components of the Dirac electronic wave 
function, with the quantum number ${\kappa}=\pm (J+\frac{1}{2})$ for
$J=l_e \mp \frac{1}{2}$ and the orbital angular momentum $l_e$.
The calculations are preformed with a realistic electronic
wave function (see \cite{Par05}). 

The magnetic dipole contribution to the hyperfine splitting
$W_{(I J)F F}$ has the form
\begin{eqnarray}
W_{(I J)F F } &=& <IJFF|\mathcal{H}|IJFF>    \\
&=& \frac{1}{2}[F(F+1)-I(I+1)-J(J+1)] a_{I}, \nonumber
\end{eqnarray}
where $a_I$ is defined as 
\begin{equation}
a_{I} = -\frac{2e \kappa \mu _{N}}{IJ(J+1)} < I I \mid \sum\limits_{i=1}^{A}
(M_{Z}^{l_i}(\mathbf{r}_i) + M_{Z}^{s_i}(\mathbf{r}_i))\mid I I >    \nonumber
\end{equation}
with the $Z$ components of the magnetic dipole moment 
$\mathbf{M}^{l_i(s_i)}(\mathbf{r}_i)$, related to the
angular momentum $l_i$ and spin $s_i$ of each nucleon; the summation runs over all the nucleons.

For an extended nuclear charge we have
\begin{eqnarray}
\mathbf{M}^{l}(\mathbf{r}_{i}) &=&g_{l}^{i}\mathbf{l}_{i} [\int \limits_{r_{i}}^{%
\infty }F^{\kappa J} G^{\kappa J}\;dr\;+\;\int \limits_{0}^{r_{i}}F^{\kappa
J} G^{\kappa J}\;(\frac{r}{r_{i}})^{3} dr] ,    \nonumber \\
\mathbf{M}^{s}(\mathbf{r}_{i}) &=&g_{s}^{i}\mathbf{s}_{i} \int \limits_{r_{i}}^{%
\infty }F^{\kappa J} G^{\kappa J}dr + \mathbf{D}%
_{i} \int \limits_{0}^{r_{i}}F^{\kappa J} G^{\kappa J}(\frac{r}{r_{i}}%
)^{3} dr ,     \nonumber
\end{eqnarray}
with 
$\mathbf{D}_{i}={-\sqrt{\frac{5}{2}}\;[\mathbf{s}^{1}\otimes C^2(%
\hat{r}_{i})]}^{1}$ and 
$C_q^{k}=\sqrt{\frac{4 \pi}{2k+1}}Y_{kq}(\hat{r}_{i})$.

The quantity $a_I$ can be expressed through the hfs 
constant for a point nucleus $a_I^{(0)}$ as
\begin{equation}
a_I \thickapprox a_I^{(0)} (1+\epsilon_{BW}+\delta)  , \label{eq10}
\end{equation}
with 
\begin{equation}
a_I^{(0)}= -\frac{2 e \kappa \;\mu _{N} \mu}{IJ(J+1)} 
\int\limits_{0}^{\infty} F_0^{\kappa J}(r) G_0^{\kappa J}(r)dr .   
\end{equation}
Here, $\mu =< I I \mid \sum\limits%
_{i=1}^{A}(g_s^i s_i + g_l^i l_i) \mid I I >$ 
defines the magnetic dipole moment of the point nucleus in nuclear
magneton units $\mu _{N}$. The functions $F_0^{\kappa J}$, 
$G_0^{\kappa J}$ are the radial parts of the Dirac electronic
wave function in the point nucleus approximation.

The hfs anomaly in the Bohr-Weisskopf effect is 
\begin{eqnarray}
\epsilon_{BW} & = & - \frac{b}{\mu}
\sum\limits_{i=1}^3 \left[ \sum\limits_{j=1}^{n_i} 
\left\{ <II| (g_s^{(j)} s_j + g_l^{(j)} l_j) K^a(r_j)|II> \right.\right.\nonumber \\
&-&  \left.\left.<II| (g_l^{(j)} l_j + D_j) K^b(r_j) |II> \right\} \right] \label{EBW},      
\end{eqnarray}
where $b=[\int_0^{\infty} F_0^{\kappa J}G_0^{\kappa J}d r]^{-1}$,
and
\begin{eqnarray}
K^a(r_j) &=& \int\limits_{0}^{r_j}F^{\kappa J}G^{\kappa J}d r  ,      \label{elwfinta} \\
K^b(r_j) &=& \int\limits_{0}^{r_j}F^{\kappa J}G^{\kappa J} (\frac{r}{r_j})^3 d r. \label{elwfintb}
\end{eqnarray}

The index $j$ is related to the $A$ nucleons; $i(1-3)$ denotes one of the 
three-clusters of $n_i$ nucleons. $g_l^{(j)}$ and $g_s^{(j)}$ 
are the gyromagnetic ratios of the $j$-th nucleon orbital motion 
and spin, respectively.

The hfs anomaly can be approximated by
\begin{eqnarray}
\epsilon_{BW}&=&- \frac{b}{\mu} 
\sum\limits_{i=1}^{3}\left[\left\langle II| (g_s^{(i)} s_i + g_l^{(i)}l_i )
K^a (r_i ) |II\right\rangle \right.   \nonumber \\
&-& \left. \left\langle II |( g_{l}^{(i)}l_i  + D_i )K^b(r_i )|II \right\rangle
\right].    \label{anom1}
\end{eqnarray}

The index $i$ is related to the three-clusters of relative 
coordinate $r_i$ and angular momentum $l_i$ in respect to the 
center of mass system, and with the appropriated expressions 
$g_s^{(i)} , g_l^{(i)}$ and $D_i$, calculated for each cluster.
The BRCS correction is defined as $\delta=1-bK^a(\infty)$ 
\cite{Ros72}. This term is defined by the nuclear charge distribution and information on the neutron distribution is contained solely in $\epsilon_{BW}$.

\section{Hfs anomaly and nuclear moments of $^9$Be in the cluster 
model \label{sec:s4}}

To calculate the magnetic dipole moment and the hfs anomaly, 
we use three systems of coordinates related to each other by 
rotation: the T-basis with the neutron as cluster $i=1$, and the 
$Y(q)$ systems ($q=1$ or $2$) with one of the $\alpha$-particles 
as cluster $i=2$ or $3$.
The index $q=1$ and $q=2$ define respectively the rotations 
$1(23)\rightarrow 2(31)$ and $1(23)\rightarrow 3(12)$ between 
the T and Y-basis. 

The transformation of the hyperspherical 
harmonic function \cite{Efr72} from the T-basis to the 
Y-basis is defined by 
\begin{equation}
\Im _{KLM_{L}}^{l_{x}l_{y}}(\Omega _{5})
=\sum\limits_{l_{x}^{\prime }l_{y}^{\prime }}
\left\langle l_{x}^{\prime }l_{y}^{\prime }
|l_{x}l_{y}\right\rangle_{KL}^{q}\Im _{KLM_{L}}^{l_{x}^{\prime }l_{y}^{\prime }}
(\Omega _{5}^{\prime }) ,
\label{transL}
\end{equation}
where $\left\langle l_{x}^{\prime }l_{y}^{\prime }
|l_{x}l_{y}\right\rangle_{KL}^{q}$ are the
Raynal-Revai coefficients \cite{Efr72}. 

The transformation of the spin part of the wave function 
is written as
\begin{equation}
\chi _{SM_{S}}^{s_1(s_2s_3)S_x}
=\sum\limits_{S_{x}^{\prime }}\left\langle S_{x}^{\prime
}|S_{x}\right\rangle _{S}^{q}
\chi _{SM_{S}}^{s_1^{\prime} (s_2^{\prime}s_3^{\prime})S_x^{\prime}}
\label{transS}
\end{equation}
with $s_1^{\prime}=s_{2(3)}$, $s_2^{\prime}=s_{3(1)}$, and 
$s_3^{\prime}=s_{1(2)}$ for $q=1(2)$,
\begin{eqnarray}
\left\langle S_{x}^{\prime
}|S_{x}\right\rangle _{S}^{q} &=& (-)^{s_{1(2)}+2s_{2(1)}+ s_3 + S_{x}^{\prime }
(S_{x})} \hat{S}_{x}\hat{S}_{x}^{\prime } \nonumber \\
 &\times& \left\{ 
\begin{array}{ccc}
s_{2(3)} & s_{3(2)} & S_{x} \\ 
s_{1} & S & S_{x}^{\prime } 
\end{array} 
\right\} ,
\end{eqnarray}
where ${\bf S=s_2+s_3+s_1}$, ${\bf S_{x}=s_2+s_3}$, and 
$ {\bf S_{x}^{\prime }=s_{3(1)}+s_{1(2)}}$.


To calculate the magnetic dipole moment, we consider the
contribution of each fragment with respect to the rest system. 
In $^9$Be, the neutron spin and the $\alpha $-particle 
orbital motion contribute to the magnetic dipole  moment as 
%
%
%
%
%
\begin{equation}
\mu =  \sum\limits_{\varsigma}\omega _{\varsigma}^{2}
\left\langle m_{s} \right\rangle g_{s}^{(n)} \label{magmom}
+ \frac{5}{9}\sum\limits_{q=1,2}\sum\limits_{\varsigma^{\prime}} \beta
_{\varsigma^{\prime}}^{2}
\left\langle m_{y}\right\rangle_q g_{l_y^{\prime}}^{(\alpha )}. 
\end{equation}
$\left\langle m_{s} \right\rangle ,\left\langle m_{y}\right\rangle_q$ 
are the expectation values of the spin and angular momentum 
projections of the fragments. $\left\langle m_{s} \right\rangle$ 
is found in the T-basis and associated with the neutron spin; 
$\left\langle m_{y}\right\rangle_q$ is found in the Y-basis 
obtained by the rotation $q=1(2)$ for $i=2(3)$ (see (\ref{transL}) 
and (\ref{transS})) and associated with the $\alpha$-particle 
orbital momentum.
In Eq. (\ref{magmom}) $\omega _{\varsigma}^{2}$  is 
the weight of the partial state in the T-basis for the channel 
with quantum numbers $\varsigma=Ll_{x}l_{y}$; and 
$\beta _{\varsigma \prime}^{2}$ is the weight of the partial 
state obtained with (\ref{transL}) and (\ref{transS})  
for each channel with quantum numbers 
$\varsigma^{\prime}=Ll_{x}^{\prime}l_{y}^{\prime}$ in the 
Y-basis. Here, we mean by channel the set $\varsigma$ of quantum 
numbers characterizing the partial waves contributing to the 
ground state wave function. 
In the case of $^9$Be represented by the system 
$\alpha + \alpha + n $, this set of quantum numbers is $L$, $l_x$, 
and $l_y$ defined in Section \ref{sec:s3}. The nuclear wave function is summed over 
the hypermomentum $K$ which is not included explicitly as the matrix 
element does not depend on it.
 
In Eq. (\ref{magmom}), 
$\mu _{n}=\frac{1}{2}g_{s}^{(n)}$ denotes the magnetic dipole moment of 
the neutron and $\mu _{l^{\prime}_y}^\alpha = 
g_{l^{\prime}_y}^{(\alpha)} m_y = 
2 \frac{A_1+A_2}{AA_3} g_{l_y}^{(p)} m_y $ is the
magnetic dipole moment of the $\alpha $-particle in the state with 
$l_{y}$ as angular momentum in the Y-basis. Note that the 
magnetic dipole moments are obtained using for the $g$ factors 
$g_{s}^{(n)}=-3.8260854(90)$ and $g_{l_y}^{(p)}=1$.
 
 

The factor $\frac{5}{9}$ in (\ref{magmom}) is
 the center of mass factor $\frac{A-A_{2(3)}}{A}$ in the 
 Y-basis corresponding to the $\alpha +(\alpha + n)$ system.
 
We assume that the $\alpha$-particle angular momentum is defined by two orbiting protons, neglecting 
the spin contribution of the nucleons.
%

The electric quadrupole  moment is 
\begin{equation}
Q = 2 \sum \limits_{i=1}^3 \left\langle II|r_i^2 C_0^{2}(\hat{r}_{i})| II \right\rangle ,
\end{equation}
where the summation runs over the fragments.

In the three-cluster model of $^9$Be, the hfs anomaly in the
Bohr-Weisskopf effect is 
found as the weighted sum
\begin{eqnarray}
\epsilon_{BW} &=& - \frac {b}{\mu_I}\left\{ 
\sum\limits_{\varsigma} \omega
_{\varsigma}^{2} \left\langle m_{s} \right\rangle g_{s}^{(n)}  
\right. \nonumber \\
& \times & \left[
K_{\varsigma}^{a}-K_{\varsigma}^{b}\left( 1\mp \frac{6}{4}
\frac{2I+1}{2(I+1)}\right) \right] \label{EpsBW} \\
&+&\left. \frac{5}{9}\sum\limits_{q = 1,2}\sum\limits_{\varsigma^{\prime}} \beta
_{\varsigma^{\prime}}^{2} \left\langle m_{y} \right\rangle_q
g_{l_y^{\prime}}^{\alpha}\left[ K_{\varsigma^{\prime}}^{a}
-K_{\varsigma^{\prime}}^{b}\right]
\right\} , \nonumber 
\end{eqnarray}
where the sign $\mp $ corresponds to $I=l_y\pm \frac{1}{2}$, and where $\mu$ is replaced by the
experimental value ($\mu_I=-1.1778(9)\mu_N$ \cite{Web76}) of the $^9$Be magnetic dipole moment.

Here we denote 
\begin{eqnarray}
K_{\varsigma}^{a} 
&=&\int\limits_{0}^{\infty }|\Phi_{\varsigma}(R)|^{2} K^a(\frac{A-A_i}{A}R)R^{2}dR \label{KA} \\
K_{\varsigma}^{b}
&=&\int\limits_{0}^{\infty }|\Phi_{\varsigma}(R)|^{2} K^b(\frac{A-A_i}{A}R)R^{2}dR  \label{KB} 
\end{eqnarray}
where the ratio $\frac{A-A_i}{A}$ takes into account the center 
of mass motion, $A_i$ is the valence fragment mass (the neutron 
mass $(i=1)$ in the T-basis, and the $\alpha $-particle 
mass $(i=2 ,3)$ in the Y-basis).

So, we can write 
\begin{equation}
\epsilon_{BW}=\sum\limits_{\varsigma} 
\omega_{\varsigma}^{2} \epsilon_{BW}^{\varsigma} \label{epschan}
\end{equation}
where 
$\epsilon_{BW}^{\varsigma}$ is obtained for each channel 
$|\varsigma >$.

\section{Results and discussion \label{sec:s5}}

\subsubsection{Three-body model of $^9$Be} 
In the calculations we use two ground
state wave functions of the $^9$Be described as the three body
$\alpha+\alpha+n$ system. These wave functions are
obtained with different $\alpha-\alpha$ and $\alpha-n$ interaction 
potentials. 

The first wave function, WF1 \cite{Priv05}, is obtained with supersymmetric 
equivalent potential \cite{Des03}. With the $\alpha $-particle
charge radius $1.676$ fm \cite{deV87} this wave function 
gives the value $2.564$ fm for the $^{9}$Be charge rms radius. This 
value agrees with the experimental ones, 
$2.519(12)$ and $2.50(9)$ fm 
(see \cite{deV87} and references therein). 
The $^{9}$Be magnetic dipole moment
$\mu _{^{9}Be}=-1.0531 \mu_N $ is less by $10\%$ compared to the
experimental values, $-1.177432(3) \mu_N $ 
\cite{Ita83} and $-1.1778(9) \mu_N $ \cite{Web76}. 
The calculated electric quadrupole  moment is 53.39 mb which is in good
agreement with the experimental value 52.88(38) mb \cite{Sun91}.

The second wave function, WF2 \cite{Gri02}, is obtained 
with the Ali-Bodmer potential \cite{Fed96} and the 
$\alpha $-neutron interaction potential \cite{Cob98}. 
The three-body interaction potential is adjusted to 
fit the three-cluster dissociation energy $1.573$ MeV.
%
The value of the $^{9}$Be charge radius, $2.707$, fm is 
overestimated compared to the experimental one. The $^{9}$Be 
magnetic dipole moment 
$\mu _{^{9}Be}= -1.316 \mu_N $ is $10\%$ larger than 
the experimental value. The electric quadrupole  moment is $Q=65.42$ mb.

\begin{table}[tbp]
\caption{\label{Table1} The weights of the partial waves 
$\protect\omega _{Ll_{x}l_{y}}^{2}$, the mean 
$^8$Be-neutron distance $r_n$, the neutron contribution 
$\epsilon_{BW}^{\varsigma (n)}$ to the values of the hfs 
anomaly obtained with WF1. 
}
\begin{ruledtabular} 
\begin{tabular}{cccc}
$\varsigma$ & $\omega_{\varsigma}^{2}$ & 
                    $r_n$  &  $\epsilon_{BW}^{\varsigma (n)}$ \\ 
$Ll_{x}l_{y}$  &   & fm    &  \%      \\ \hline
101    &  0.498355 & 3.274 & -0.0332 \\ 
121    &  0.315975 & 3.159 & -0.0229 \\ 
221    &  0.152790 & 2.976 &  0.0123 \\ 
123    &  0.019661 & 3.758 & -0.0312 \\ 
223    &  0.003674 & 3.585 & -0.0286 \\ 
143    &  0.006608 & 3.689 & -0.0303 \\ 
243    &  0.001595 & 4.397 &  0.0245 \\ 
\end{tabular}
\end{ruledtabular}
\end{table} 
 
\begin{table}[tbp]
\caption{\label{Table2} The same as in Table \ref{Table1}
calculated with the wave function WF2.}
\begin{ruledtabular} 
\begin{tabular}{cccc}
$\varsigma$ & $\omega _{\varsigma}^{2}$ & 
                    $r_n$  & $\epsilon_{BW}^{\varsigma (n)}$ \\ 
$Ll_{x}l_{y}$  &   & fm    &  \%     \\ \hline
101    &  0.592607 & 3.778 & -0.0402 \\ 
121    &  0.286695 & 3.381 & -0.0248 \\ 
221    &  0.090244 & 3.284 &  0.0141 \\ 
123    &  0.020740 & 3.870 & -0.0313 \\ 
223    &  0.002710 & 3.753 & -0.0296 \\ 
143    &  0.005362 & 3.770 & -0.0299 \\ 
243    &  0.001154 & 3.760 &  0.0179 \\ 
\end{tabular}
\end{ruledtabular}
\end{table}

The difference between these wave functions is in the radial $y$
dependence of the valence neutron wave function (obtained by 
integration over the $x$ coordinate and the angles), 
the core charge radius and the weights of the partial states 
$L$, $l_x$, $l_y$ for the ground state wave function 
(see Tables \ref{Table1} and \ref{Table2}).

Tables \ref{Table1} and 
\ref{Table2} show  the neutron contribution
$\epsilon_{BW}^{\varsigma(n)}$ 
to the hfs anomaly (first term in (\ref{EpsBW}))
and the root mean square distance of
the valence neutron $r_n$ from the $^8$Be center of mass
in each channel $\varsigma$. 

The contributions to the hfs anomaly in the Bohr-Weisskopf effect, 
$\epsilon_{BW}^{(n)}$ and $\epsilon_{BW}^{(\alpha)}$,
and the BRCS correction $\delta$ calculated with
WF1 and WF2 are listed in Table \ref{Table3}.



The contribution of the hfs anomaly from the $\alpha$-particle
orbital motion, $\epsilon_{BW}^{(\alpha)}$ (second term in (\ref{EpsBW}))
is small compared to that from the neutron, $\epsilon_{BW}^{(n)}$; 
and its variation is small also (see Table \ref{Table3}).

\begin{table}[tbp]
\caption{\label{Table3} The $^{9}$Be charge rms radius, the neutron radial distance $r_n$, the contributions to the hfs anomaly from the
neutron spin ($\epsilon_{BW}^{(n)}$) and from the
$\alpha$-particle orbital motion ($\epsilon_{BW}^{(\alpha)}$), 
the values of $\epsilon_{BW}$ and of 
$\delta$ calculated for the two wave functions $WF1,2$ ($\alpha$-particle radius: $1.676$ fm)
are compared to the (core+neutron) results. WF2$^*$ refers to another choice of the $\alpha$-particle radius ($1.46$ fm).}

\begin{ruledtabular} 
\begin{tabular}{lcccc}
Value                           &core+neutron & WF1       & WF2      & WF2$^*$    \\ \hline
rms (fm)                        & 2.519       & 2.564     & 2.707    & 2.533   \\ 
$r_n$ (fm)                      &             & 3.207     & 3.621    &         \\ 
$\epsilon_{BW}^{(n)}$ (\%)      &             & -0.02281  & -0.03059 &         \\ 
$\epsilon_{BW}^{(\alpha)}$ (\%) &             & 0.00085   &  0.00088 &          \\ 
$\epsilon_{BW}$ (\%)            &  -0.0236    & -0.02112  & -0.02882 & -0.03032  \\ 
$\delta$                        &  -0.0451    & -0.04644  & -0.04926 & -0.04634  \\ 
$\epsilon$                      &  -0.0687    & -0.06756  & -0.07809 & -0.07666 \\ 
\end{tabular}
\end{ruledtabular}
\end{table}

To explore the sensitivity of the results to the charge radius of 
the $\alpha$-clusters in $^9$Be, we vary the value of 
the $\alpha$-particle charge radius from $1.676$ to $1.636$ fm.
The last value of the charge radius is obtained with regard to
the negative contribution of the neutron charge distribution, $(0.34)^2$ fm$^2$
(see \cite{Sic05,Ana99}).

\begin{figure}
\includegraphics{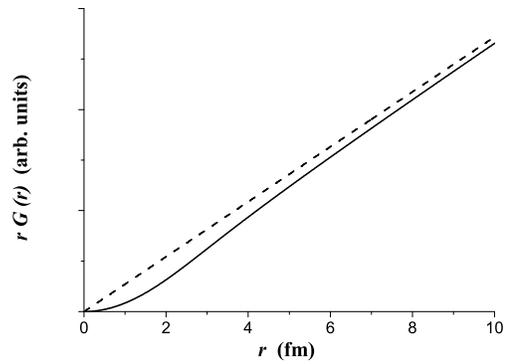}
\caption{\label{fig1} The electronic wave functions
obtained for the distributed nuclear charge, $G$ (dashed line),
and for the point nuclear charge, $G_0$ (solid line).}
\end{figure}

Correspondingly, the $^9$Be charge radius changes from 2.564 to 
2.534 fm for WF1 and from 2.707 to 2.678 fm for WF2.
Owing to the radial behavior of the electronic wave functions
(see Fig. \ref{fig1}) entering the expressions (\ref{elwfinta}),
(\ref{elwfintb}), and (\ref{EpsBW})-(\ref{KB}), with a smaller 
charge radius (rms and rms$_C$) we get a larger value of the 
hfs anomaly in the Bohr-Weisskopf effect. Thus, in our case, 
we get an increase in the hfs anomaly value 
$\epsilon_{BW}$ by 1.6\% for WF1 and 0.5\% for WF2.

The $\alpha$-$\alpha$ distance is smaller when calculated with WF1, so
correspondingly, the $^9$Be charge radius is
smaller and the value of the BRCS correction $\delta$ is 
less (see Table \ref{Table3}).
$\delta$ varies within 6\% depending on the description of the nuclear wave function.


$\epsilon_{BW}$ and the total hfs anomaly 
$\epsilon$ which is the 
sum ($\epsilon_{BW}^{(n)}+\epsilon_{BW}^{(\alpha)}+\delta$), 
are mainly determined by the neutron distribution. 
There is no direct correspondence of the hfs anomaly
value to the nuclear charge radius.
For example, if we put the radius of the $\alpha$-particle at
1.46 fm, the $^9$Be charge radius obtained with WF2 is 
2.533 fm and the hfs anomaly $\epsilon$=-0.07666\% (see Table \ref{Table3} WF2$^*$).
These values are larger than those obtained with WF1 with the
charge radius 2.564 fm. 
Thus, even with smaller charge radius values one can
get larger values of $\epsilon_{BW}$ and $\epsilon$.

Therefore, we can conclude that the hfs anomaly is more sensitive 
to the neutron spatial distribution than to the charge 
distribution of the whole nucleus, and
the value $r_n$ is a crucial parameter  
for the hfs value in the Bohr-Weisskopf effect. 

With the different nuclear wave functions, the $r_n$ value varies of about
$10\%$ and the hfs anomaly $\epsilon_{BW}$ of about 30\%. 
As the value of the hfs anomaly $\epsilon_{BW}$ is 
about two times smaller than $\delta$,the total
hfs anomaly $\epsilon$ varies within 14\% with the choice of the wave function.


We should also mention that
the value of the hfs anomaly is very sensitive to the
contribution of the different partial states in 
the $^{9}$Be ground state wave function. 
In particular, the hfs anomaly $\epsilon_{BW}^{(n)}$ in the channels
$\varsigma=121$ and $\varsigma=221$ (that is of 50\% of the
nuclear wave function) is twice as small as $\epsilon_{BW}^{(n)}$
in the channel $\varsigma=101$. Thus, the relative weights of 
these states are of major importance. As found in \cite{Vor95} the magnetic dipole moment in the
$\alpha+\alpha+n$ cluster model is also rather sensitive to
these weights. In the case of $^{9}$Be, three channels mostly contribute 
to the magnetic dipole moment and the hfs anomaly, $\varsigma=101$, 
$\varsigma=121$, and $\varsigma=221$. The magnetic dipole moment 
in this case is \cite{Vor95}
\begin{equation}
\mu \thickapprox -1.857 \omega_{101}^2- 1.191 \omega_{121}^2
+ 1.914 \omega_{221}^2 + \mu _{res}.
\end{equation}

According to \cite{Vor95} the experimental magnetic dipole 
moment value can be reproduced under the condition
$\omega_{221}^2 < 16$ \%. Otherwise the 
calculated value is underestimated.

Let us analyze the correlation between the
weights and the values of the $^{9}$Be hfs anomaly,  
magnetic dipole and electric quadrupole moments, noting that the analysis is model dependent.

\begin{figure}
\includegraphics{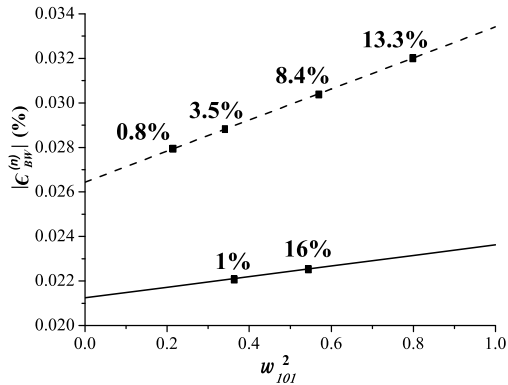}
\caption{\label{fig2} The variation of the Bohr-Weisskopf hfs anomaly value 
from the neutrons with the weight $\omega_{101}^{2}$ for WF1 (solid line) and WF2 
(dashed line)($\alpha$-particle radius: $1.676$ fm). The dots on the lines are the values in agreement 
with the electric quadrupole  moment(see text). The weight $\omega_{221}^{2}$ is indicated for each point.} 
\end{figure}

To estimate how the hfs anomaly $\epsilon_{BW}^{(n)}$
varies with the weights of the different states, we consider 
the contributions of these three channels only
(so that $\omega_{101}^2+\omega_{121}^2 + \omega_{221}^2 + \omega_{res}^2=1$)
and find the weights satisfying the experimental value of the
$^9$Be magnetic dipole moment ($\mu_I=-1.177432(3)\mu_N$). Under
this condition, we get the hfs anomaly value plotted in 
Fig. \ref{fig2} as a function of $\omega_{101}^2$ for the wave 
functions WF1 and WF2 (solid and dashed lines, respectively).

Similarly one can 
express the electric quadrupole  moment through the weights of 
these dominant states. For WF1 we get
\begin{eqnarray}
Q & \thickapprox & 3.157 \omega_{101}^2 - 4.553 \omega_{121}^2 - 
4.527 \omega_{221}^2 \nonumber \\ 
 &+&  2 \omega_{101} \omega_{121} 32.859  
+ 2 \omega_{101} \omega_{221} 32.039 + \label{quad1} \\
 &+&  2 \omega_{121} \omega_{221} 7.314 + Q_{res},  \nonumber
\end{eqnarray}
 where $Q_{res}=11.97$ mb.
 
For WF2 this relation is
\begin{eqnarray}
Q & \thickapprox & 4.204 \omega_{101}^2 - 6.635 \omega_{121}^2 - 
6.662  \omega_{221}^2 \nonumber \\ 
&+& 2 \omega_{101} \omega_{121} 41.908 
+ 2 \omega_{101} \omega_{221} 41.610 \label{quad2} \\
&+& 2 \omega_{121} \omega_{221} 9.918 +Q_{res}, \nonumber
\end{eqnarray}
where $Q_{res}=13.11$ mb.

$\omega_{res}^2$, $\mu _{res}$ and $Q_{res}$ are the contributions of the neglected channels respectively
to the weights, the magnetic dipole and electric quadrupole moments. 
The square dots on the lines in Fig. \ref{fig2} point 
the hfs values obtained with the weights which also
satisfy the experimental value of the $^9$Be electric quadrupole  
moment 52.88(38) mb.
For each wave function we get a few
points - two for WF1 or even four for WF2. 
On this figure we also report the corresponding values of the weight 
$\omega_{221}^2$ ($\omega_{221}^2<16\%$).
One can see the ranges of the hfs anomaly values  
obtained with WF1 and WF2 respectively 
Therefore, an experimental estimation 
of the hfs anomaly in the Bohr-Weisskopf effect could give
values for the weights of the partial states in the
ground state wave function.


The figure shows that the hfs anomaly is a very critical 
quantity to test the nuclear wave function, all other
parameters being equivalently well described, 
in particular, the electronic part. 

\subsubsection{Comparison with the core+neutron model \label{sec:s6}}

In Ref. \cite{Par05}, the hfs anomaly for $^{9}$Be was calculated 
within the core+neutron model of $^{9}$Be. 
The $^9$Be ground state wave function was given by the 
superposition of states 
\begin{eqnarray}
\left| ^{9}{\rm Be} \left( 3/2^{-}\right) \right\rangle 
&=&\omega_{0^+} \left|
[^{8}Be \left( 0^{+}\right) \otimes n_{p_{3/2}}]_{3/2^{-}}\right\rangle  \label{WF9}  \\
&+&\omega_{2^+} \left| [^{8}Be \left( 2^{+}\right) \otimes
n_{p_{3/2}}]_{3/2^{-}}\right\rangle  .    \nonumber   
\end{eqnarray}
corresponding to the $^8$Be core 
in the ground (0$^{+}$)  and excited (2$^{+}$) states with
the neutron separation energies 1.665 and 4.705 MeV, respectively. 

With the weights $\omega_{0^+}^2= 0.535$ and $\omega_{2^+}^2=0.465$ 
obtained with the spectroscopic factors from Ref. \cite{Coh67} the 
magnetic dipole moment is $\mu=-1.0687$ $\mu_N$ and $\epsilon_{BW}$, $\delta$ and $\epsilon$ have the values reported in Table \ref{Table3}. 
The $\epsilon_{BW}$ is close to the values $\epsilon_{BW}=-0.0249$\% \cite{Fuj99} (obtained with the
weights $\omega_{0^+}^2=\omega_{2^+}^2=0.5$) and 
$\epsilon_{BW} =-0.0243$\% from \cite{Yam00}. 

The $\epsilon_{BW}$ values obtained for the $0^+$ state in the 
(core+neutron) model ($\epsilon_{BW}= - 0.0440$\%) and for the 
$l_x = 0$ state of $WF1$ or $WF2$ in the ($\alpha + \alpha + n$) 
model ($\epsilon_{BW}= -0.0332\%$ or  $-0.0402$\%) are relatively 
close to each other. On the contrary, the $\epsilon_{BW}$ 
values for the different partial states $l_x = 2$ in the 
three cluster model exceed by a few times 
(see Tables \ref{Table1} and 
\ref{Table2}) the value obtained for the $2^+$ state
in the (core + neutron) model ($\epsilon_{BW}= -0.0063$\%).


The BRCS correction obtained 
with the two-body wave function
is $-0.0451$\%, close to that obtained in the three-body 
calculations. Thus in the core+neutron model we get 
$\epsilon=-0.0687$\%, to be compared with 
the values $\epsilon=-0.06756$\% and $\epsilon=-0.07809$\% 
obtained with the three clusters 
$ \alpha + \alpha + n $ wave functions.


Thus the clustering effect, revealing itself in the set 
of states contributing to the ground state wave function,
lead to a variation of the hfs value $\epsilon$ of less than 2\% for WF1 and of about 
14\% for WF2. Compared to the  $^{11}$Be nucleus the difference in the
 value of the hfs anomaly in the Be isotopes is about 25\%. 
This value gives us the accuracy of the measurements
of the hfs anomaly needed to study clustering effects
in light nuclei.


This results corroborates the 
conclusion in Ref. \cite{Fuj99}, that the
value of the hfs anomaly reflects the extended neutron 
distribution in $^{11}$Be and might indicate a neutron halo, 
but the difference for the different isotopes is not so pronounced 
as was found in Ref. \cite{Fuj99}.

\section{Conclusion \label{sec:s7}}

In the present paper, we have calculated the hfs 
anomaly in the $^{9}$Be$^+$ ion with the nucleus described in a 
three-cluster model. The $1s^22s$ electronic wave 
function is obtained taking into account the charge distribution of the clustered ($\alpha+\alpha +n$)nucleus and the shielding effect of two electrons in the $1s^2$ configuration. 


The result of the calculations strongly depends 
on the weights of the partial waves contributing to the ground state
wave function. Together with the magnetic dipole and electric quadrupole moments 
the value of the hfs anomaly can be used to study the
clustering effects in neutron rich light nuclei.

The total hfs anomaly is the sum of $\delta$ and $\epsilon_{BW}$.
The BRCS correction $\delta$ is only determined
by the nuclear charge distribution and slightly varies from
isotope to isotope. The value of the BRCS correction is comparable 
or larger than the value of $\epsilon_{BW}$. 
The hfs anomaly in $^{11}$Be differs from that in $^{9}$Be by 25\%. 
The clustering effect leads to variations of the hfs value
within 15\%. The calculated magnitude and differential change in the value of the hfs 
anomaly is indicative of the experimental precision that must be achieved to study 
the clustering effect and the neutron distribution in neutron rich light nuclei.


\section{Acknowledgments}

The authors are grateful to Dr P. Descouvemont and Dr. L.V. Grigorenko 
for helpful discussions and for having provided the numerical values
of the wave functions they have calculated for $^9$Be in a three-body model.

This paper has been supported by the Belgian Program P5-07 of Inter-university Attraction Poles initiated by the Belgian-state Federal Services for Scientific Politics. Y.L.P. has received an one year postdoctoral research grant from the Cooperation of the Belgian Scientific Politics with Central and Oriental European countries.

\end{document}